\documentclass[11pt]{article}
\usepackage{latexsym,epic,eepic}
\usepackage{amsmath,amsfonts}
\usepackage{epsfig}

\newtheorem{theorem}{Theorem}[section]
\newtheorem{corollary}[theorem]{Corollary}
\newtheorem{lemma}[theorem]{Lemma}

\newcommand{\qed}{\hfill$\diamond$}
\newcommand{\pf}{{\bf Proof: }}

\begin{document}

\title{The Dichotomy of List Homomorphisms for Digraphs}

\author{Pavol Hell \thanks{Simon Fraser University, Burnaby, B.C., Canada V5A 1S6} \ \ and \ \ Arash Rafiey 
\thanks{Simon Fraser University and IDSIA, Supported by Swiss National Science Foundation project N.200020-122110/1  Approximation Algorithms for Machine Scheduling Through
Theory and Experiments III} 
}

\date{}

\newcommand{\eopf}{\raisebox{0.8ex}{\framebox{}}}
\newenvironment{proof}%
{\noindent{\bf Proof.}\ }%
{\hfill\eopf\par\bigskip}%

\maketitle

\begin{abstract}
The Dichotomy Conjecture for constraint satisfaction problems 
has been verified for conservative problems (or, equivalently, 
for list homomorphism problems) by Andrei Bulatov. An earlier
case of this dichotomy, for list homomorphisms to undirected 
graphs, came with an elegant structural distinction between the 
tractable and intractable cases. Such structural characterization 
is absent in Bulatov's classification, and Bulatov asked whether
one can be found. We provide an answer in the case of digraphs;
the technique will apply in a broader context. 

The key concept we introduce is that of a digraph asteroidal triple 
(DAT). The dichotomy then takes the following form. If a digraph 
$H$ has a DAT, then the list homomorphism problem for $H$ is 
NP-complete; and a DAT-free digraph $H$ has a polynomial time 
solvable list homomorphism problem. DAT-free graphs can be 
recognized in polynomial time.
\end{abstract}

\section{Introduction}

The framework of constraint satisfaction problems (CSP's) allows
a unification of many natural problems arising in applied computer
science and artificial intelligence. In recent years, it has also become
central in theoretical computer science, with most of the interest 
driven by the Dichotomy Conjecture formulated by T. Feder and M.
Vardi in \cite{fv}.

A general constraint satisfaction problem consists of a set of variables
with values in a common domain, and a set of constraints limiting the 
values the variables can take. The theoretical investigations frequently 
focus on the so-called {\em non-uniform} CSP's, where the constraints
are restricted by a certain finite {\em template}. The Dichotomy Conjecture
simply says that for each such template the corresponding non-uniform
problem is polynomial or NP-complete. Two original motivating examples
for the Dichotomy Conjecture were Schaeffer's dichotomy classification
of Boolean satisfiability problems \cite{schae}, and Hell-Ne\v set\v ril's
dichotomy of graph homomorphism problems \cite{hn}. The first case 
corresponds to the templates in which the common domain has only 
two values (say 0, 1); the second case corresponds to templates which
are undirected graphs. Since that time, a number of other special 
cases have been established, e.g., \cite{barto,kozik,bula,f,fms}, including
the case of conservative problems \cite{bul} discussed below. In most of
these cases, progress has been made possible by an algebraic approach
pioneered by Jeavons, Cohen, and Gyssens \cite{jcg}. In particular, Bulatov, 
Jeavons, and Krokhin \cite{bjk} have established that the complexity of 
a non-uniform CSP only depends on the so-called polymorphisms of the 
template. This fundamentally affected the quest for the Dichotomy Conjecture,
and in particular allowed more concrete statements of the expected distinction
between tractable and intractable cases \cite{bjk,larose,maroti}, cf. the survey
\cite{ccc}.

The special case of conservative CSP's is one of the early successes of the 
algebraic method; here dichotomy has been settled by Bulatov \cite{bul}.
In his paper, Bulatov notes that his dichotomy lacks the combinatorial
insights offered by earlier special case of undirected graphs \cite{fhh2}.
We provide such combinatorial insights in the case of directed graphs.
This yields the first polynomial time distinction between the tractable and
the intractable cases for conservative CSP's in the case of digraphs. In the 
process, we also simplify Bulatov's classification in terms of polymorphisms.

Our technique is a combination of forbidden structure characterizations
typical of structural graph theory, and the polymorphism approach typical of 
the algebraic method. It will extend to more general templates, at least to 
those corresponding to several binary relations.

\section{Preliminaries}

In this paper we mostly focus on templates that are digraphs.
A {\em digraph} $H$ is a finite set $V(H)$ of {\em vertices}, together with a 
binary relation $E(H)$ on the set $V(H)$; the elements of $E(H)$ are called 
{\em arcs} of $H$. A {\em homomorphism} of a digraph $G$ to a digraph $H$ 
is a mapping $f : V(G) \rightarrow V(H)$ which preserves arcs, i.e., such that
$uv \in E(G)$ implies $f(u)f(v) \in E(H)$. The CSP with template $H$, also
known as the {\em homomorphism problem for $H$} is the decision problem
in which the instance is a digraph $G$ and the question is whether or not $G$
admits a homomorphism to $H$. We note that we view an undirected graph 
$H$ as a special case of a directed graph, in which the relation $E(H)$ is
symmetric.

More generally, a {\em relational structure} $H$ is a finite set $V(H)$, and a
sequence of relations $E_1(H), E_2(H), \dots, E_t(H)$, where each relation
$E_i(H)$ has a finite arity $a_i$. The sequence $a_1, a_2, \dots, a_t$ is 
called the {\em type} of $H$. If $G$ is a relational structure of the same 
type as $H$, a homomorphism of $G$ to $H$ is a mapping 
$f : V(G) \rightarrow V(H)$ which preserves all relations.
Feder and Vardi \cite{fv} have pioneered the view of non-uniform CSP's 
as homomorphism problems for a template $H$ that is a relational structure. 
They have also identified the special case when $H$ is a digraph (that is,
$k=1, a_1=2$) as crucial for the Dichotomy Conjecture - if the conjecture 
holds for templates that are digraphs, then it holds in general \cite{fv}.

We say that a non-uniform constraint satisfaction problem is {\em conservative} 
if the template $H$ is a relational structure that contains all possible $2^{|V(H)|-1}$ 
non-empty unary relations on $V(H)$. Conservative CSP's can be equivalently 
described in terms of list homomorphisms \cite{bul,fh,fhh,fhh2} as follows.
For a fixed template relational structure $H$, the {\em list homomorphism 
problem to $H$}, denoted $LHOM(H)$, asks whether or not an input relational
structure $G$, of the same type as $H$, equipped with lists $L(v), v \in V(G)$, 
admits a homomorphism $f : G \rightarrow H$, such that for each $v \in V(G)$
we have $f(v) \in L(v)$.

The problem LHOM$(H)$ has been thoroughly studied for templates $H$ that
are undirected graphs \cite{fh,fhh,fhh2}. For example, for {\em reflexive} graphs 
$H$ (every vertex has a loop), the problem LHOM$(H)$ is polynomial time solvable 
if $H$ is an interval graph, and is NP-complete otherwise \cite{fh}. For {\em 
irreflexive} graphs (no vertex has a loop), the problem LHOM$(H)$ is polynomial 
time solvable if $H$ is a bipartite graph whose complement is a circular arc graph, 
and is NP-complete otherwise \cite{fhh}. For general graphs, where some vertices 
may have loops and others don't, there is also a structural, albeit somewhat more
technical, distinction  \cite{fhh2}. The problem LHOM$(H)$ has also been studied 
when $H$ is a reflexive digraph, in \cite{catarina,adjust}.

These results have motivated a focus on conservative CSP's, leading to a full proof 
by Bulatov \cite{bul} of the dichotomy of LHOM($H$) for all templates $H$. To explain
the results, we first give the necessary definitions concerning polymorphisms.

Let $H$ be a relational structure and $k$ a positive integer. An mapping $f : V(H)^k
\rightarrow V(H)$ is a {\em polymorphism} of $H$, of order $k$, if it is compatible
with all relations $E_i(H)$. (The mapping $f$ is compatible with the relation $E_i(H)$
if $(u^1_1,u^1_2,\dots,u^1_{a_i}), \dots, (u^k_1,u^k_2,\dots,u^k_{a_i}) \in E_i(H)$
implies $(f(u^1_1,\dots,u^k_1),\dots,f(u^1_{a_i},\dots,u^k_{a_i})) \in E_i(H)$.)
A polymorphism $f$ is {\em conservative} if $f(u_1,u_2,\dots,u_k)$ 
always is one of $u_1, u_2, \dots, u_k$. A polymorphism $f$ of order two is {\em 
commutative} if $f(u,v)=f(v,u)$ for all $u, v$, and is {\em associative} if 
$f(u,f(v,w))=f(f(u,v),w)$ for all $u, v, w$. A commutative and
associative polymorphism is called a {\em semi-lattice}. It is easy to see that a
conservative semi-lattice polymorphism $f$ on $H$ defines a linear ordering $<$ 
on $V(H)$ such that $f(u,v) = \min(u,v)$. (It is enough to set $u < v$ exactly when
$f(u,v)=u$.) Conversely, if $f(u,v) = \min(u,v)$ is a polymorphism, then this polymorphism 
is clearly conservative, commutative and associative; we call it a {\em min-ordering}. In
other words, $<$ is a min-ordering of $H$ just if it satisfies the following property:
if $uv \in E(H)$ and $u'v' \in E(H)$, then $\min(u,u')\min(v,v') \in E(H)$. A polymorphism
$f$ of order three is a {\em majority function} if $f(u,u,v)=f(u,v,u)=f(v,u,u)=u$ for any
$u$ and $v$. A polymorphism $f$ of order three is called {\em Maltsev} if $f(u,u,v)=f(v,u,u)=v$ 
for any $u$ and $v$. It is known that if $H$ admits a conservative majority function or 
a conservative Maltsev polymorphism, or a conservative semi-lattice polymorphism 
(e.g. a min-ordering), then the problem $LHOM(H)$ is polynomial time solvable 
\cite{bjk,fv,welzl}. Bulatov's result states that, locally, these are the only reasons 
for the polynomiality of LHOM$(H)$.

\begin{theorem}\cite{bul}\label{bulatov}
Let $H$ be a relational structure. If, for each pair of vertices $u, v$ of $H$, there exists 
a conservative polymorphism $f_{u,v}$ of $H$, that either is binary and $f | _{u,v}$ is 
a semi-lattice, or is ternary and $f | _{u,v}$ is majority, or Maltsev, then LHOM$(H)$ is 
polynomial time solvable.

Otherwise, LHOM$(H)$ is NP-complete.
\end{theorem}

This proves dichotomy of LHOM($H$) (and hence for all conservative CSP's), and provides
a criterion for distinguishing the tractable and intractable cases. As noted in \cite{bul},
the criterion is polynomial in $H$ provided $V(H)$ is fixed. It is also pointed out in \cite{bul}
that the criterion lacks the combinatorial elegance and the structural information of the earlier
results for undirected graphs.

We provide a simpler classification in the case when the template $H$ is a digraph.
The characterization is similar to the spirit of the earlier combinatorial classifications
for graphs \cite{fh,fhh,fhh2}, and gives structural information about obstructions that 
cause intractability. It provides for digraphs the first criterion that is polynomial in $V(H)$. 
As a byproduct, we also prove that in the case of digraphs, the statement of Bulatov's 
theorem can be simplified as follows.

\begin{corollary}\label{bulatovlike}
Let $H$ be a digraph. If, for each pair of vertices $u, v$ of $H$, there exists 
a conservative polymorphism $f_{u,v}$ of $H$, that either is binary and $f | _{u,v}$ is 
a semi-lattice, or is ternary and $f | _{u,v}$ is majority, then LHOM$(H)$ is 
polynomial time solvable.

Otherwise, LHOM$(H)$ is NP-complete.
\end{corollary}

We have recently learned that A. Kazda \cite{kaz} has proved that if a digraph $H$ 
admits a Maltsev polymorphism, it must also admit a majority polymorphism. Corollary 
\ref{bulatovlike} can be viewed as complementing  Kazda's result, by showing that 
for conservative polymorphisms of digraphs, Maltsev polymorphisms are not 
needed even locally.

\section{Asteroidal Triples}

Recall that for reflexive graphs $H$, the problem LHOM($H$) is polynomial
time solvable if $H$ is an interval graph and is NP-complete otherwise \cite{fh}.
According to the theorem of Lekkerkerker and Boland \cite{lekbol}, a graph is 
an interval graph if and only if it does contain an induced cycle of length at least 
four, or an {\em asteroidal triple}, i.e., three vertices $a, b, c$ any two of which 
are joined by a path avoiding the neighbours of the third vertex. Since induced 
cycles of length at least six are easily seen to contain asteroidal triples themselves, 
we may view asteroidal triples as the principal structures in undirected graphs
$H$ that cause the NP-completeness of LHOM$(H)$. We will introduce a digraph 
relative of an asteroidal triple that we call a {\em digraph asteroidal triple}, or DAT.
Even though its somewhat technical definition makes a DAT only a distant 
relative of the simple concept of an asteroidal triple, DATs play a pivot role for 
digraphs as asteroidal triples (together with induced four- and five-cycles) play 
for undirected graphs - namely they are the only obstructions to polynomiality 
of the problem LHOM($H$), cf. Theorem \ref{main}.

Let $H$ be a digraph. We say that $uv \in E(H)$ is a {\em forward arc} of $H$
(or just an {\em arc} of $H$); in that case we also say that $vu$ is a {\em 
backward arc} of $H$. We define two walks $P = x_0, x_1, \dots, x_n$ and 
$Q = y_0, y_1, \dots, y_n$ in $H$ to be {\em congruent}, if they follow the 
same pattern of forward and backward arcs. Specifically, by this we mean 
that $x_ix_{i+1}$ is a {\em forward arc} (respectively {\em backward arc})
of $H$ if and only if $y_iy_{i+1}$ is a forward (respectively backward) arc 
of $H$. If $P$ and $Q$ as above are congruent walks in $H$, we say that 
$P$ {\em avoids} $Q$, if there is no arc $x_iy_{i+1}$ in the same direction 
(forward or backward) as $x_ix_{i+1}$.

Note that a walk (or path) has a beginning and an end. A {\em reversal} of
a walk $P = x_0, x_1, \dots, x_n$ is the walk $P = x_n, x_{n-1}, \dots, x_0$.

\vspace{3mm}

An {\em invertible pair} in $H$ is a pair of vertices $u, v$, such that

\begin{itemize}
\item
there exist congruent walks $P$ from $u$ to $v$ and $Q$ from $v$ to $u$,
such that $P$ avoids $Q$,

\item
and there exist congruent walks $P'$ from $v$ to $u$ and $Q'$ from $u$ to $v$,
such that $P'$ avoids $Q'$.
\end{itemize}

Note that it is possible that $P'$ is the reversal of $P$ and $Q'$ is the reversal
of $Q$, as long as both $P$ avoids $Q$ and $Q$ avoids $P$.

Let $H$ be a digraph. We introduce the following auxiliary digraph $H^+$. The 
vertices of $H^+$ are all ordered pairs $(u,v)$, where $u, v$ are vertices of $H$.
There is an arc from $(u,v)$ to $(u',v')$ in $H^+$ in one of the following situations:

\begin{enumerate}
\item
$uu' \in E(H), vv' \in E(H), uv' \not\in E(H)$, or
\item
$u'u \in E(H), v'v \in E(H), v'u \not\in E(H)$.
\end{enumerate}

In case 1, we say the arc from $(u,v)$ to $(u',v')$ is a $+$ arc in $H^+$, in case 2, 
we say it is a $-$ arc. Two directed walks in $H^+$ are {\em similar} if they have 
the same pattern of $+$ and $-$ arcs (thus, in particular, the same length).

\vspace{3mm}

We make the following observations:

\begin{enumerate}
\item
$u, v$ is an invertible pair if and only if $(u,v), (v,u)$ are in the same strong component
of $H^+$;
\item
$H^+$ contains an arc from $(u,v)$ to $(u',v')$ if and only if it contains an arc from
$(v',u')$ to $(v,u)$;
\item
if $u, v$ is an invertible pair in $H$ and $(p,q)$ is in the same strong component of $H^+$
as $(u,v)$, then $p, q$ is also an invertible pair in $H$.
\end{enumerate}

The observation 2 will be called the {\em skew-symmetry} of $H^+$. It is useful for
proving many facts about $H^+$, including the observation 3 above.

A {\em permutable triple} in $H$ is a triple of vertices $u, v, w$ together with six 
vertices $s(u), b(u)$, $s(v), b(v)$, $s(w), b(w)$, which satisfy the following condition.

\begin{itemize}
\item
For any vertex $x$ from $u, v, w$, there exists a walk $P(x,s(x))$ from $x$ to $s(x)$ 
and two walks $P(y,b(x))$ (from $y$ to $b(x)$), and $P(z,b(x))$ (from $z$ to $b(x)$),
congruent to $P(x,s(x))$, such that $P(x,s(x))$ avoids both $P(y,b(x))$ and $P(z,b(x))$.
\end{itemize}

Note that since $P(x,s(x))$ avoids both $P(y,b(x))$, and $P(z,b(x))$, a permutable
triple yields two similar directed walks $P(x;y), P(x;z)$ from $(x,y)$ and $(x,z)$ to 
$(s(x),b(x))$ in $H^+$.

Recall that an undirected asteroidal triple $u, v, w$ is defined by the property that for any 
vertex $x$ from $u, v, w$, there exists a walk joining the other two vertices which is 
avoided by the neighbours of $x$. Our definition of a permutable triple already sounds 
vaguely reminiscent of this. However, we will need another technical condition.

A {\em digraph asteroidal triple} (DAT) is a permutable triple in which each of the 
three pairs $(s(u),b(u)),$ $(s(v),b(v))$, and $(s(w),b(w))$ is invertible. This turns 
out to imply that the entire permutable triple is inside one strong component of
$H^+$. 

\begin{theorem}\label{all_in}
If $u, v, w$ is a DAT, then there exist, for each permutation $x, y, z$ of $u, v, w$, 
walks $P(x,s(x))$, $P(y,b(x))$, and $P(z,b(x))$, as in the definition above, such 
that all the associated walks $P(x;y), P(x;z)$ lie entirely inside one fixed strong 
component $C$ of $H^+$.

In particular, all six pairs $(u,v), (v,u),$ $(u,w),$ $(w,u), (v,w),$ $(w,v)$, and all six 
pairs $(s(u),b(u)),$ $(s(v),b(v))$, $(s(w),b(w))$ are invertible, and belong to $C$.
\end{theorem}

\pf
Indeed, consider in $H$ the three vertices $u, v, w$ of a DAT, the three invertible pairs 
$s(u), b(u)$, and $s(v), b(v)$, and $s(w), b(w)$, and the nine walks $P(u,s(u))$, $P(v,b(u))$, 
$P(w,b(u))$, $P(v,s(v))$, $P(u,b(v))$, $P(w,b(v))$, and $P(w,s(w))$, $P(u,b(w))$, 
$P(v,b(w))$, from the definition. Consider now the walks $P(u;v), P(u;w)$, and $P(v;u)$,
$P(v;w)$, and $P(w;u), P(w,v)$, joining $(u,v), (u,w)$ to $(s(u),b(u))$, and $(v,u), (v,w)$ to 
$(s(v),b(v))$, and $(w,u), (w,v)$ to $(s(w),b(w))$, respectively, in $H^+$.

Suppose $x$ is any vertex from $u, v, w$. Since $s(x), b(x)$ is an invertible pair, 
we also have in $H^+$ a directed walk $Q(x)$ from$(s(x),b(x)$ to $(b(x),s(x))$.
Consider now the following directed walk from $(u,v)$ to $(v,u)$: concatenate $P(u;v)$
from $(u,v)$ to $(s(u),b(u))$, with $Q(u)$ from $(s(u),b(u))$ to $(b(u),s(u))$, and
then concatenated with the skew-symmetric walk to $P(u;v)$ (taking us from 
$(b(u),s(u))$ to $(v,u)$. Replacing the last segment by the skew-symmetric walk
to $P(u;w)$ yields a directed walk from $(u,v)$ to $(w,u)$. By similar concatenations 
we see that all the pairs $(u,v), (v,u),$ $(u,w),$ $(w,u),$ $(v,w),$ $(w,v)$, as well as 
all vertices on the walks $P(u;v), P(u;w)$, $P(v;u)$, $P(v;w)$, $P(w;u), P(w,v)$, including 
$(s(u),b(u)),$ $(s(v),b(v))$, $(s(w),b(w))$ are in the same component of $H^+$.
\qed

Thus we can equivalently define a DAT as a permutable triple in a self-coupled 
strong component of $H^+$.

We note, for future reference, that for the proof we only used walks from 
$(b(u),s(u))$ to $(s(u),b(u))$, from $(b(v),s(v))$ to $(s(v),b(v))$, and from 
$(b(w),s(w))$ to $(s(w),b(w))$. (Although invertibility of these pairs also 
ensures directed walks from $(b(u),s(u))$ to $(s(u),b(u))$ and so on, we 
did not use these walks in the proof.)

\vspace{3mm}

We will prove the following classification.

\begin{theorem}\label{main}
Let $H$ be a digraph.

If $H$ contains a DAT, the problem LHOM$(H)$ is NP-complete.

If $H$ is DAT-free, the problem LHOM$(H)$ is polynomial time solvable.
\end{theorem}

Deciding whether or not a given digraph $H$ contains a DAT is easily
seen to be polynomial in the size of $V(H)$. One only needs to check 
for connectivity properties in suitable auxiliary digraph defined on the 
triples of vertices of $H$. Specifically, let $H^{++}$ be the digraph with 
the vertex set $V(H)^3$ and an arc from $(u,v,w)$ to $(u',v',w')$ just
when $H$ has arcs $uu', vv', ww'$ but not $uv'$ and not $uw'$, or
$H$ has an arcs $u'u, v'v, w'w$ but not $v'u$ and not $w'u$. Then
$u, v, w$ is a DAT if and only if for every permutation $x, y, z$ of 
$u, v, w$, the digraph $H^{++}$ contains an invertible pair $s, b$ 
such that $(s,b,b)$ is reachable from $(x,y,z)$.

\section{The Dichotomy}

We first prove the following fact.

\begin{theorem}\label{dat}
If $H$ contains a DAT, then LHOM($H$) is NP-complete.
\end{theorem}

\pf
Consider the three vertices $u, v, w$, the three invertible pairs 
$(s(u), b(u))$, $(s(v), b(v))$, $(s(w), b(w))$,
the three congruent walks $P(u,s(u))$, $P(v,b(u))$, $P(w,b(u))$, the 
three congruent walks $P(v,s(v))$, $P(u,b(v))$, 
$P(w,b(v))$, and the three congruent walks $P(w,s(w))$, $P(u,b(w))$, 
$P(v,b(w))$, from the definition of a DAT. 
Consider also the six walks arising from the fact that $(s(u), b(u))$, 
$(s(v), b(v))$, $(s(w), b(w))$ are invertible pairs - 
namely a walk $P(s(u),b(u))$ (from $s(u)$ to $b(u)$), that avoids 
a congruent walk $P(b(u),s(u))$ (from $b(u)$ to $s(u)$), 
and the correspondingly defined walks $P(s(v),b(v)),$ $P(b(v),s(v))$, 
$P(s(w),b(w)),$ and $P(b(w),s(w))$.

According to Theorem \ref{all_in} all the pairs $(u,v), (v,u),$
$(u,w)$, $(w,u), (v,w),$ $(w,v)$, $(s(u),b(u)),$ $(b(u),s(u)),$ $(s(v),b(v)),$ 
$(b(v),s(v)),$ $(s(w),b(w)),$ $(b(w),s(w))$ belong to the same strong 
component $C$ of $H^+$. We first consider the case that $C$ is 
{\em symmetric}, in the sense that for each arc $(a,b)(c,d)$ in $C$,
the reversal $(c,d)(a,b)$ is also an arc of $C$. Observe that this means 
that pairs in $C$ can be joined by walks which avoid each other in both 
directions, and in particular, we may assume that both $P(v,b(u))$ 
and $P(w,b(u))$ avoid $P(u,s(u))$, etc., and that $P(b(u),s(u))$ avoids 
$P(s(u),b(u))$, and so on.

In this case we shall use gadgets called "choosers", as defined in the proof 
of \cite{hombook} Lemma 5.3.6. We repeat the definition here, applied to 
digraphs. Let $i, j$ be distinct vertices from $u, v, w$ and let $I, J$ be subsets 
of  $\{u, v, w\}$. A {\em chooser} $Ch(i,I;j,J)$ is a digraph $X$ with specified 
vertices $x$ and $y$, and with lists $L(p) \subseteq V(H)$, for $p \in V(X)$, 
such that any list homomorphism $f$ of $X$ to $H$ has $f(x)=i$ and $f(y) \in I$ 
or $f(x)=j$ and $f(y) \in J$; and for any $i' \in I$ and $j' \in J$ there is a list 
homomorphism $f$ of $X$ to $H$ with $f(x)=i$ and $f(y)=i'$ and a list 
homomorphism $g$ of $X$ to $H$ with $g(x)=j$ and $g(y)=j'$.

According to the proof of Lemma 5.3.6 in \cite{hombook}, if there exist 
choosers $Ch(i,\{i,k\};$ $j,\{j,k\})$ and $Ch(i,\{i\}; j,\{k\})$, for any permutation 
$ijk$ of $uvw$, then LHOM($H$) is NP-complete. (Those proofs are stated 
in terms of undirected graphs $H$, but they apply verbatim to digraph choosers 
as defined here.) We construct the choosers as follows. Suppose $i=u$, $j=v, k=w$ 
(the other cases are similar), and consider a path $P$ congruent to the walk 
$P(u,s(u))$ concatenated with its reversal. Then define $Ch(u,\{u\};v,\{w\})$ to 
be $P$ with $x$ being the initial vertex of $P$ and $y$ being the terminal vertex 
of $P$, and with the lists $L(x)=\{u,v\}$, $L(y)=\{u,w\}$, and for all other vertices 
$p$ of $P$ the list $L(p)$ consists of the corresponding vertices on $P(u,s(u))$ 
concatenated with its reversal, and on $P(v,b(u))$ concatenated with the reversal 
of $P(w,b(u))$. This ensures that $Ch(u,\{u\};v,\{w\})$ satisfies the definition of a
chooser as required.

Finally, to define $Ch(u,\{u,w\};v,\{v,w\})$, we take four vertices $a, b, c, d$, and 
place one copy of $P$ between $a$ and $c$ (identifying $a$ with $u$ and $c$ 
with $v$), and another copy of $P$ between $b$ and $d$ (identifying in a similar 
manner). Then we define $P'$ as a path congruent to the walk $P(v,s(v))$ 
concatenated with its reversal, and place a copy of $P'$ between $c$ and $b$ 
and another copy of $P'$ between $d$ and $a$. It can be checked that the 
resulting digraph satisfies the conditions for a chooser $Ch(u,\{u,w\};v,\{v,w\})$
with the specified vertices $x=a$ and $y=b$. This completes the proof of 
NP-completeness of LHOM$(H)$ when $C$ is symmetric.

Otherwise, let $(a,b)(c,d)$ be an arc in the strong component $C$ such 
that $(c,d)(a,b)$ is not an arc of $H^+$. We now claim that we can choose 
the congruent walks $P(s(u),b(u))$ and $P(b(u),s(u)),$ so that $P(s(u),b(u))$ 
avoids $P(b(u),s(u))$ (as required by their definition), but $P(b(u),s(u))$ does 
not avoid $P(s(u),b(u))$. Indeed, in the strong component $C$ there is a 
directed walk from $(s(u),b(u))$ to $(a,b)$ and from $(c,d)$ to $(b(u),s(u))$, 
which together with the arc from $(a,b)$ to $(c,d)$ yields the desired walks 
$P(s(u),b(u))$ and $P(b(u),s(u))$. A similar statement holds for 
$P(s(v),b(v)), P(b(v),s(v))$, and $P(s(w),b(w)), P(b(w),s(w))$.

Let $Q_w$ be a path congruent to the walk $P(u,b(w))$ concatenated with 
$P(b(w),s(w))$ and then with the reversal of $P(w,s(w))$. Note that $Q_w$ 
admits a homomorphism to $H$ with the initial vertex of $Q_w$ mapping to 
$u$ or to $v$ and the terminal vertex of $Q_w$ mapping to $w$, as well as
a homomorphism mapping the initial vertex to $w$ and the terminal vertex to 
either $u$ or $v$, using the fact that $P(s(w),b(w)),$ and $P(b(w),s(w))$ are 
congruent walks. Since the walk $P(b(w),s(w))$ does not avoid the walk 
$P(s(w),b(w))$, the portion of $Q_w$ corresponding to $P(b(w),s(w))$ 
admits a homomorphism to $H$ taking the first and last vertex to $b(w)$, 
whence $Q_a$ also admits a homomorphism taking its initial vertex to 
$u$ or $v$ and its terminal vertex also to $u$ or $v$. In summary, $Q_a$ 
admits a homomorphism to $H$ taking its initial and terminal vertices to 
$u, v, w$ in any combination, except possibly both to $w$. We now introduce 
lists on the vertices of $Q_a$ which limit mapping each vertex of $Q_a$ to 
any of the images implicit in the homomorphisms used above (mapping the 
initial and terminal vertex to $u, v, w$ in all possible combinations), including 
the lists $\{u, v, w\}$ for the initial and terminal vertices. We now claim there 
is no list homomorphism (consistent with these lists) which takes both the 
initial and the terminal vertex of $Q_a$ to $w$. This follows from the fact 
that $P(w,s(w))$ avoids both $P(u,b(w))$ and $P(v,b(w))$, that $P(s(w),b(w))$ 
avoids $P(b(w),s(w))$, and that the reversal of both $P(u,b(w))$ and $P(v,b(w))$ 
avoid the reversal of $P(w,s(w))$, by the skew-symmetry of $H^+$.

Since the walks $P(b(u),s(u))$ (respectively $P(b(v),s(v))$) do not avoid the 
walks $P(s(u),b(u))$ (respectively $P(s(v),b(v))$), we may similarly construct 
paths $Q_v, Q_u$ with the property that there exists a homomorphism of $Q_v$ 
(respectively $Q_u$) to $H$ taking the initial vertex to $x \in \{u,v,w\}$ and the 
terminal vertex to $y \in \{u,v,w\}$ if and only if $x$ and $y$ are not both equal 
to $v$ (respectively $u$). Finally, we let $Q$ be the digraph obtained from the 
union of $Q_u, Q_v, Q_w$ by identifying all three initial vertices into one, and 
identifying all three terminal vertices into one. It follows that $Q$ admits a 
homomorphism to $H$ taking the (combined) initial vertex to $x \in \{u,v,w\}$ 
and the (combined) terminal vertex to $y \in \{u,v,w\}$ if and only if $x$ and $y$ 
are not equal. By a standard trick, this allows the reduction of 3-COL to LHOM$(H)$. 
(Given an undirected graph $G$, we replace each edge by a copy of $Q$; the 
resulting graph admits a list homomorphism to $H$ if and only if $G$ is 3-colourable.)

This proves the NP-completeness of LHOM$(H)$ when $C$ is not symmetric.
\qed

On the other hand, we now proceed to show that a DAT-free
digraph $H$ has a tractable LHOM($H$). We will use Theorem
\ref{bulatov}. The following result, although not needed in the 
proof of Theorem \ref{main}, illustrates our approach, and
introduces a technique used in the proof of Theorem \ref{twopols}.
In the journal version we will also provide a similar characterization
of digraphs that admit a conservative semi-lattice (or, equivalently,
a min-ordering). We note that this last result applies only in digraphs,
and the existence of conservative semi-lattice is NP-complete even
for two binary relations \cite{arnauld}.

\begin{theorem}\label{weaktriple}
A digraph $H$ admits a conservative majority function if and only 
if it has no permutable triple.
\end{theorem}

\pf 
Assume that there is no permutable triple in $H$. We proceed to define a 
conservative majority function $\mu$ on $H$ as follows. Consider three 
vertices $u, v, w$ . Let $x$ be one vertex of $u, v, w$, and $y, z$ the other 
two vertices. We say that $x$ is a {\em distinguisher} for $x, y, z$, if for any 
three mutually congruent walks from $x, y, z$ to $s(x), b(x), b(x)$ respectively, 
the first walk does not avoid one of the other two walks. Since there is no 
permutable triple, at least one of $u, v, w$ must be a distinguisher for $u, v, w$. 
This definition can be applied even if the vertices $u, v, w$ are not distinct. Note 
that if, say, $u=v$ then no walk starting in $u$ can avoid a walk starting in $v$, 
whence $u$ is a distinguisher, and similarly for $v, w$.

We define the values of $\mu$ as follows: for a triple $(u,v,w)$, we set 
$\mu(u,v,w)$ to be the first vertex from $u, v, w$, in this order, that is a 
distinguisher for $u, v, w$. Note that the last remark ensures that if $u=v$
or $u=w$, then $\mu(u,v,w)=u$. On the other hand, if $v=w$, it is possible
that $u, u\neq v$ is a distinguisher of $u, v, w$, and we make an exception
and define $\mu(u,v,w)=v$.

It remains to show $\mu$ is a polymorphism. Thus, for contradiction, 
suppose $uu' \in E(H)$, $vv' \in E(H), ww' \in E(H)$, 
and $\mu(u,v,w)\mu(u',v',w') \not\in E(H)$. (A symmetric proof applies if 
$u'u \in E(H)$, $v'v \in E(H), w'w \in E(H)$, and $\mu(u',v',w')\mu(u,v,w) 
\not\in E(H)$.) This clearly implies that at least one of the triples $u, v, w$ 
or $u', v', w'$ are distinct vertices. Suppose one of the triples, say $u, v, w$ 
has a repetition. If $u=v$, then $\mu(u,v,w)=u$, and $\mu(u',v',w')=w'$ 
(since $\mu(u,v,w)\mu(u',v',w') \not\in E(H)$). This contradicts the fact that 
$w'$ is a distinguisher, since the walk $w'w$ avoids both paths $u'u, v'w$. 
A similar proof applies if $u=w$. If $v=w$, we have defined $\mu(u,v,w)=v$
regardless of whether $u$ is a distingiusher, so we have the same proof
as well. It remains to consider the case when both triples $x, y, z$ and 
$x', y', z'$ consist of distinct vertices. Assume first
that $\mu(x,y,z)=x, \mu(x',y',z')=y'$. Note that $xz' \not\in E(H)$, else $y'$
would not be a distinguisher of $x', y', z'$, because of the paths $y'y, x'x,
z'x$. The fact that $\mu(x',y',z')=y'$ means that $x'$ is not a distinguisher 
for $x', y', z'$, thus there exist congruent walks $X', Y', Z'$ from $x', y', z'$ 
respectively, such that $Y', Z'$ end at the same vertex, and $X'$ avoids 
$Y'$ and $Z'$. Then $xx'$ followed by $X'$, together with $yy'$ followed 
by $Y'$ and $zz'$ followed by $Z'$ are also congruent walks, and the walk
$xx', X'$ avoids the other two walks, contradicting the fact that $x$ is a
distinguisher for $x, y, z$.

Because of the symmetry between $x, y, z$ and $x', y', z'$, the only other 
cases to consider are $\mu(x,y,z)=x, \mu(x',y',z')=z'$ and $\mu(x,y,z)=y, 
\mu(x',y',z')=z'$; in both situations a similar proof applies.

For the converse, let $\mu$ be a majority function on $H$, and assume that $u, v, w$ 
is a permutable triple on $H$ with $\mu(u,v,w)=u$. There exist mutually congruent 
walks from $u, v, w$ to $s(u), b(u), b(u)$ respectively, where the first walk avoids 
the second two walks. Let $u_1, v_1, w_1$ be the first vertices on these walks
(respectively), just after $u, v, w$. Then we must have $\mu(u_1,v_1,w_1)=u_1$,
since $u$ lacks the right kind of arc to $v_1$ and $w_1$. Similarly, for the $i$-th 
vertices of these walks, we must have $\mu(u_i,v_i,w_i)=u_i$; this is impossible 
as it would imply that $\mu(s(u),b(u),b(u))=s(u)$. Symmetric arguments handle the 
cases when $\mu(u,v,w)=v$ and $\mu(u,v,w)=w$. Thus there is no conservative 
majority function.
\qed

Specifically, we shall show that a DAT-free digraph $H$ admits two special
conservative polymorphisms - a binary polymorphism $f$ and a ternary polymorphism
$g$ - such that for all pairs $u, v$ of vertices of $H$, the restriction $f | _{u,v}$ is a 
semi-lattice, or the restriction $g | _{u,v}$ is a majority. It will then follow from Theorem 
\ref{bulatov} that for DAT-free digraphs $H$, the problem LHOM$(H)$ is polynomial time 
solvable. It will also follow, using also Theorem \ref{dat}, that we may omit the mention
of Maltsev polymorphisms from the statement of Theorem \ref{bulatov}, to obtain
Corollary \ref{bulatovlike}.

\begin{theorem}\label{twopols}
Suppose $H$ is a DAT-free digraph.

Then $H$ admits a binary polymorphism $f$ and a ternary polymorphism $g$
such that 

\begin{itemize}
\item
if $u, v$ is not invertible then $f | _{u,v}$ is semi-lattice, and
\item
if $u, v$ is invertible then $g | _{u,v}$ is majority.
\end{itemize}

\end{theorem}

\begin{corollary}\label{polo}
If $H$ is DAT-free, then LHOM$(H)$ is polynomial time solvable.
\qed
\end{corollary}

\pf
The rest of this section is devoted to the proof of Theorem \ref{twopols}.
Thus we shall assume for the remainder of the section that $H$ is a
DAT-free digraph. We will first define the binary polymorphism $f$.
To start, we define $f(x,x)=x$ for all vertices $x$. It remains to define
$f$ on pairs $(x,y)$ that are vertices of $H^+$.

We say that the pair $(x,y)$ is a {\em special} vertex of $H^+$ if there is in $H^+$
a directed path from $(x,y)$ to $(y,x)$. Note that skew-symmetry of $H^+$ implies
that if $(x,y)$ has a directed path to a special vertex, then $(x,y)$ is also special.
In particular, a strong component either has all its vertices special (in which case
we say it is a {\em special component}), or none of its vertices are special. For 
each component $C$ of $H^+$ we define the {\em coupled} component 
$C' = \{ (u,v) : (v,u) \in C \}$. We say that $C$ is {\em co-special} if $C'$ is
special. It is possible that $C=C'$, in which case we say that $C$ is {\em 
self-coupled}. Note that a strong component $C$ is self-coupled if and only
if it is both special and co-special; this happens if and only if all pairs in $C$
are invertible.

The {\em condensation} of a digraph $H$ is obtained from $H$ by identifying
each strong component of $H$ to a vertex and placing an arc between the
shrunk vertices just if there was an arc between the corresponding strong
components. The condensation of any digraph $H$ is acyclic. The condensation
of $H^+$ has a particular structure, arising from the properties of special and
co-special strong components. Recall that if $C_1$ is special and reachable
from $C_2$, then $C_2$ is also special; thus by skew-symmetry, if $C_1$ is
co-special and can reach $C_2$, then $C_2$ is also co-special. This means
that following any directed path in the condensation of $H^+$ we first encounter
some (possibly none) special strong components, followed by at most one 
self-coupled strong component, and then by the co-special strong components
corresponding to the initial special strong components, in the reverse order.

We make the following obvious but important observation; most of the proofs 
make a repeated use of it.

\begin{lemma}
Suppose $f$ is any binary polymorphism of $H$.

If $(x,y)$ dominates $(x',y')$ in $H^+$, then $f(x,y)=x$ implies $f(x',y')=y'$.
\qed
\end{lemma}

This suggests the following definition. If $(x,y)$ lies in a co-special strong component 
which is not self-coupled, we set $f(x,y)=x$. If $(x,y)$ lies in a special strong component 
which is not self-coupled, we set $f(x,y)=y$. For all pairs $(x,y)$ in self-coupled strong 
components, we set $f(x,y)=y$. It remains to define $f(x,y)$ for pairs $(x,y)$ in strong 
components that are neither special nor co-special. 

Consider now the subgraph $X$ of the condensation of $H^+$ induced by vertices 
corresponding to strong components that are neither special not co-special. It
is easy to see that this subgraph has a topological sort, i.e., a linear ordering of
its vertices (strong components of $H^+$), $C_1, C_2, \dots C_k, C_{k+1}, \dots, C_{2k}$
such that any arc of $H^+$ goes from some $C_i$ to some $C_j$ with $i < j$, and 
such that each $C_i$, $i \leq k$, is coupled with the corresponding $C_{2k+1-i}$.
(To see this, set $C_{2k}$ be any vertex of out-degree zero in $X$, and let $C_1$
be the strong component of $H^+$ coupled to $C_{2k}$. Then remove $C_1$ and
$C_{2k}$ from $X$ and repeat.) Now we set $f(x,y)=y$ for all strong components 
corresponding to $v_1, v_2, \dots v_k$ and $f(x,y)=x$ for all $(x,y)$ in strong 
components $C_{k+1}, \dots C_{2k}$, and $f(x,y)=y$ for all $(x,y)$ in strong 
components $C_1, C_2, \dots C_k$. 

This defines a mapping of $V(H)^2$ to $V(H)$. The definition ensures that if
$f(x,y)=x$, then $f(x',y')=x'$ for any $(x',y')$ dominated by $(x,y)$.

It is easy to see that $f$ is a polymorphism; indeed, suppose if $xx', yy'$ are 
arcs of $H$ and $f(x,y)f(x',y')$ is not an arc of $H$. Asume $f(x,y)=x, f(x',y')=y'$. 
(The other case $f(x,y)=y, f(x',y')=x'$ is similar.) Then there is an arc in $H^+$ from 
$(x,y)$ to $(x',y')$, contradicting the property of $f$ mentioned just above.

It also follows from the definition that $f(x,y)=f(y,x)$ for all pairs $(x,y)$ in special
but not co-special strong components, co-special but not special strong components,
and all strong components that are neither special not co-special. In other words, 
$f(x,y)=f(y,x)$ holds unless $(x,y)$ is in a self-coupled component, i.e., unless
$x, y$ is an invertible pair. 

Thus $f$ is a conservative polymorphism of $H$ which is commutative on pairs that 
are not invertible. It follows that it is semi-lattice on those pairs. (For conservative 
polymorphisms, associativity for pairs follows easily from commutativity.)

\begin{lemma}\label{f}
The mapping $f$ is a polymorphism of $H$, and is semi-lattice on all pairs $(x,y)$ 
that are not invertible.
\qed
\end{lemma}

We now proceed to define the ternary conservative polymorphism $g$. It also
depends on the location of the pairs of arguments in the  condensation of $H^+$, 
but it will be more convenient to define it in terms of the polymorphism $f$ (which 
itself was defined in terms of the condensation of $H^+$). 

The proof has some similarity to that of Theorem \ref{weaktriple}. 
Consider three vertices $u, v, w$ . Let $x$ be one vertex of $u, v, w$, and 
$y, z$ the other two vertices. We say that $x$ is a {\em weak distinguisher} 
for $x, y, z$, if for any three mutually congruent walks from $x, y, z$ to 
$s(x), b(x), b(x)$ respectively, {\em such that} $s(x), b(x)$ {\em is an invertible 
pair}, the first walk does not avoid one of the other two walks. Since there is no 
DAT, at least one of $u, v, w$ must be a weak distinguisher for $u, v, w$.

Given a triple $(x,y,z)$ of vertices of $H$, we consider the six-tuple of values 
$$G(x,y,z) = [f(x,y), f(y,x), f(y,z), f(z,y), f(x,z), f(z,x)],$$ each value being a vertex 
$x, y,$ or $z$. Note that each value can occur at most four times, for instance $z$ 
can not occur in the first or second coordinate of $G(x,y,z)$. Moreover, our definition 
of $f$ ensures that if $f(x,y) \neq f(y,x)$, then $f(x,y)=y, f(y,x)=x$. We set $g(x,y,z)$ 
to be the value which occurs most frequently in the six-tuple $G(x,y,z)$.
If there is a tie, we choose as $g(x,y,z)$ the first vertex amongst the tied vertices, 
in the order of preference, first $x,$ then $y,$ then $z$, that is a weak distinguisher 
for $x, y, z$. (The detailed consideration of cases below shows such a weak 
distinguisher always exists.)

Note that if $x$ fails to be a weak distinguisher of $x, y, z$, then the pairs
$(x,y)$ and $(x,z)$ have similar directed walks in $H^+$ to some $(s(x),b(x))$
that lies in a self-coupled strong component. It follows from our definition
of $f$ that this implies that $f(x,y)=y, f(x,z)=z$. This is helpful is deciding
whether or not $x, y,$ or $z$ can fail to be a weak distinguisher. In particular,
it implies that the value $g(x,y,z)$ chosen is always a weak distinguisher of 
$x, y, z$.

\begin{lemma}\label{g}
The mapping $g$ is a polymorphism of $H$, and is majority on all 
pairs $(x,y)$ that are invertible.
\qed
\end{lemma}

\pf
For ease of reference, we list here all the possible six-tuples $G(x,y,z)$ and
the resulting $g(x,y,z)$. (In the first case, we grouped several possibilities 
together by using "?" as a wild card.)

\begin{enumerate}
\item
$[x,x,?,?,x,x] \rightarrow x, [y,y,y,y,?,?] \rightarrow y, [?,?,z,z,z,z] \rightarrow z$
\item
$[x,x,z,z,z,x] \rightarrow x, [y,x,y,y,x,x] \rightarrow x, [y,y,z,y,z,z] \rightarrow y$
\item
$[x,x,y,y,z,x] \rightarrow x, [x,x,z,y,z,z] \rightarrow z, [y,y,z,y,x,x] \rightarrow y$

$[y,y,z,z,z,x] \rightarrow z, [y,x,y,y,z,z] \rightarrow y, [y,x,z,z,x,x] \rightarrow x$
\item
$[y,x,z,y,x,x] \rightarrow x, [y,x,z,y,z,z] \rightarrow z, [x,x,z,y,z,x] \rightarrow x$

$[y,y,z,y,z,x] \rightarrow y, [y,x,y,y,z,x] \rightarrow y, [y,x,z,z,z,x] \rightarrow z$
\item
$[x,x,y,y,z,z] \rightarrow x, [y,y,z,z,x,x] \rightarrow x$
\item
$[y,x,z,y,z,x]$. In this case any of $x, y, z$ could fail to be a weak distinguisher;
since each occurs twice in the six-tuple, $g(x,y,z)$ is just the first weak
distinguisher of $x, y, z$, in the order $x, y, z$.
\end{enumerate}

The claim about being majority on invertible pairs follows directly from the
definition of $g$. In fact, the six-tuples $G(x,x,y)$, $G(x,y,x)$, and $G(y,x,x)$ 
each contain two values $f(x,x)=x$, and in case of invertible pairs $x, y$
the other four values are evenly divided between $x$ and $y$. Therefore
$g(x,x,y)=g(x,y,x)=g(y,x,x)=x$ if $x, y$ is an invertible pair. (We observe in
passing, although we do not need it, that in fact $g(x,x,y)=g(x,y,x)=g(y,x,x)=x$ 
on all pairs $x, y$ except those where $f(x,y)=f(y,x)=y$.)

We proceed to prove that $g$ is a polymorphism of $H$. Thus we consider
two triples $(x,y,z), (x',y',z')$ where $xx', yy', zz'$ are arcs of $H$, and show
that $g(x,y,z)g(x',y',z')$ is also an arc of $H$. If instead $x'x, y'y, z'z$ are 
arcs of $H$, then $g(x',y',z')g(x,y,z)$ is also an arc of $H$, by a proof that
is literally the same, except it reverses all arcs listed here as going from the 
unprimed to the primed vertices. (In other words, the arcs we consider here
as forward arcs are viewed as backward arcs.) We proceed by contradiction, 
and assume that $g(x,y,z)g(x',y',z')$ is not an arc of $H$. The proof is technical, 
but the arguments in most cases are very similar. Hence we will focus here only 
on the case when $g(x,y,z)=x, g(x',y',z')=y'$. (By symmetry, this covers also the
case when $g(x,y,z)=y, g(x',y',z')=x'$. The other cases, checked by very similar 
arguments, are $g(x,y,z)=x, g(x',y',z')=z'$, and $g(x,y,z)=y, g(x',y',z')=z'$.)

Thus we shall assume throughout this proof that the pair $(x,y)$ dominates the pair 
$(x',y')$ in $H^+$. 

If four of the values in $G(x,y,z)$ are $x$, and four of the values in $G(x',y',z')$ are 
$y'$, i.e., if both six-tuples are in case (1), then in particular $f(x,y)=x, f(x',y')=y'$. 
Since $(x,y)$ dominates $(x',y')$ in $H^+$, this contradicts the fact that $f$ is a 
polymorphism.

This argument still applies if four of the values in $G(x,y,z)$ 
are $x$ and three of the values in $G(x',y',z')$ are $y'$, i.e., 
if the six-tuple of primed vertices is in cases (2, 3, 4),
because we must either have $f(x,y)=x,$ $f(x',y')=y'$ or $f(y,x)=x,$ 
$f(y',x')=y'$. The latter case similarly contradicts the fact that $f$ is 
a polymorphism, since skew-symmetry implies that $(y',x')$ dominates 
$(y,x)$ in $H^+$.

In the case three of the values in $G(x,y,z)$ are $x$ and three of the values 
in $G(x',y',z')$ are $y'$, i.e., both six-tuples are in the cases (2, 3, 4), similar 
arguments handle all situations except when $f(y,x)=x, f(x,y)=y,$ and 
$f(x',y')=y', f(y',x')=x'$. In this situation, we would have $f(x,z)=f(z,x)=x$, 
and $f(y',z')=f(z',y')=y'$. We must not have the arc $xz'$ in $H$, else 
$(y',z')$ would dominate $(y,x)$ while $f(y',z')=y', f(y,x)=x$. Now $(x,z)$ 
dominates $(x',z')$ and hence $f(x',z')=f(z',x')=x'$. This means that 
$G(x',y',z')$ contains three $x'$ and three $y'$, and since both are weak 
distinguishers, we should have had $g(x',y',z')=x'$.

This leaves us to consider the cases where at least one of the triples 
$x, y, z$ or $x', y', z'$, say the first one, has two occurrences of each
$x, y, z$. These are the situations in cases (5, 6), i.e., $[x,x,y,y,z,z]$, 
$[y,y,z,z,x,x]$, and $[y,x,z,y,z,x]$.

We continue to assume that $g(x,y,z)=x$ and $g(x',y',z')=y'$ and $xy'$ 
is not an arc of $H$; recall that the pair $(x,y)$ dominates the pair 
$(x',y')$ in $H^+$. 

In case $G(x,y,z) = [x,x,y,y,z,z]$, from $f(x,y)=x$ we obtain $f(x',y')=x'$ which 
further implies that $f(y',x')=x'$. Now there are at most two values $y'$ in the 
six-tuple $G(x',y',z')$, and at least two values $x'$. Since $f(x',y')=x'$, the 
vertex $x'$ is a weak distinguisher of $x', y', z'$, and this contradicts the 
definition of $g(x',y',z')$.

In case $G(x,y,z) = [y,y,z,z,x,x]$, we first note that $f(z',y')=z'$ implies $f(y',z')=y'$, 
and so $G(x',y',z')$ has more $z'$ than $y'$, or there are at least as many $x'$
as $y'$ and $x'$ is a weak distinguisher of $x', y', z'$. This contradicts the 
definition of $g(x',y',z')=y'$. Thus we may assume that $f(z',y')=y'$, whence 
$zy'$ must be an edge of $H$ (else $f(z,y)=z$ contradicts $f(z',y')=y'$). 
Therefore $(x,z)$ dominates $(x',y')$ in $H^+$ and hence $f(x',y')=x'$ 
and thus $f(y',x')=x'$. It again follows that either $z'$ occurs more frequently 
than $y'$, or $x'$ is a weak distinguisher and occurs at least as frequently as $y'$, 
contradicting the definition of $g(x',y',z')=y'$.

Finally, in case of $[y,x,z,y,z,x]$, as $(y',x')$ dominates $(y,x)$ in $H^+$,
we have $f(y,x)=x$ imply $f(y',x')=x'$. It is readily checked from (1-6) that 
now only the following cases from (3, 4, 6) lead to $g(x',y',z')=y'$:

\begin{itemize}
\item
$[y',x',y',y',z',z']$
\item
$[y',x',y',y',z',x']$
\item
$[y',x',z',y',z',x']$
\end{itemize}

In the first case, $[y',x',y',y',z',z']$, we have $f(z',x')=z'$ and $f(z,x)=x$, thus
$xz'$ must be an edge of $H$. Now there are congruent walks (with one arc)
in $H$ from $y', x', z'$ to $y, x, x$ respectively, such that the first avoids the
other two and $x, y$ is an invertible pair ($f$ is non-commutative only on 
invertible pairs); this means that $y'$ is not a weak distinguisher of $x', y', z'$ 
and contradicts the choice of $g(x',y',z')=y'$.

In the second case, $[y',x',y',y',z',x']$, similarly, we must have the arc $zy'$
in $H$, and so we again have congruent walks (with one arc) in $H$ from 
$x, y, z$ to $x', y', y'$ respectively, such that the first avoids the other two
and $x', y'$ is an invertible pair, implying that $x$ is not a weak distinguisher
of $x, y, z$, and contradicting the assumption that $g(x,y,z)=x$.

In the last case, we have all pairs in $x, y, z$ and in $x', y', z'$ invertible.
Since $x'$ is not a weak distinguisher of $x', y', z'$, but $x$ is a weak
distinguisher of $x, y, z$, we must have in $H$ the arc $xz'$; this arc
yields congruent (one-arc) walks in $H$ from $y', x', z'$ to $y, x, x$, and
$x, y$ is an invertible pair. This again contradicts the choice of $g(x',y',z')=y'$.
\qed

Theorem \ref{twopols} now follows from Lemmas \ref{f} and \ref{g}.

\section{Conclusions and Future Directions}

In the journal version of the paper we shall offer several applications
of Theorem \ref{main}, classifying the complexity of problems LHOM($H$)
for digraphs $H$ restricted to some natural classes of digraphs. We
illustrate these results with the following two examples. In these cases, it 
can be shown that either a conservative semi-lattice, or a conservative majority
polymorphism, suffice to cover the tractable cases. A similar phenomenon
for reflexive digraphs is reported in \cite{catarina} - conservative
semi-lattice polymorphisms suffice to cover the tractable cases, as 
conjectured in \cite{ccc} Conjecture 5.5.

\begin{theorem}
Let $H$ be a digraph whose underlying graph is a tree.

If $H$ has a semi-lattice polymorphism (or, equivalently, a min-ordering), 
then $LHOM(H)$ is polynomial time solvable.

Otherwise $H$ contains a DAT and LHOM($H$) is NP-complete.
\end{theorem}

\begin{theorem}\label{cyc}
Let $H$ be a digraph whose underlying graph is a cycle. 

If $H$ has a conservative majority function, then LHOM($H$) is 
polynomial time solvable. 

Otherwise $H$ contains a DAT and LHOM($H$) 
is NP-complete.
\qed
\end{theorem}

Feder \cite{f} studied the complexity of the (non-list) homomorphism problem 
for digraphs $H$ whose underlying graph is a cycle. The situation for list
homomorphisms turned out to be significantly simpler; we will provide a
concrete description of the tractable cases.

\vspace{3mm}

We close with a few questions.

\begin{enumerate}
\item
Investigate the class of DAT-free digraphs.
\item
Find a more efficient algorithm to recognize if a digraph is DAT-free.
\item
Find a simpler polynomial time algorithm for LHOM($H$) when $H$ 
is a DAT-free digraph.
\item
Find other types of relational structures $H$ for which there is a
polynomial time distinction between the tractable and intractable 
cases of LHOM($H$).
\end{enumerate}

Regarding 1, 2, we point out that in the undirected case the class of 
AT-free graphs is quite popular, as it unifies several known graph 
classes, has interesting structural properties, and allows efficient 
algorithms for computational problems 
intractable in general  \cite{hajo,chang,dgc,dgc1}. The recognition 
problem for AT-free graphs is easily seen to be polynomial, but the 
search for really efficient recognition of AT-free graphs appears to 
be continuing \cite{eki}. For 3, we mean an algorithm simpler than
that inherent in \cite{bul}. For 4, we note that our techniques will 
extend to relational structures containing only binary relations: 
one can make the obvious modifications in the definition of $H^+$, 
making it a relational structure of the same type as $H$, and proceed
quite analogously. We will include more details in the journal version.

{\bf Acknowledgements}

We thank Tom\' as Feder, Xuding Zhu, and Zden\v ek Dvo\v r\' ak
for many useful conversations related to these results.

We also thank NSERC Canada and IRMACS SFU for financial and 
logistic support, without which this research could not have occurred.

\end{document}